\documentclass[aps,prb,twocolumn,notitlepage]{revtex4-2}

\usepackage{lipsum}
\usepackage{amsmath,graphicx,color}
\usepackage{bm}
\usepackage{natbib}
\usepackage{placeins}
\setcitestyle{square}
\usepackage{braket}
\usepackage{makerobust}
\usepackage[dvipsnames]{xcolor}
\usepackage{ulem}

\begin{document}

\title{Implementation of Topological Quantum Gates in Magnet-Superconductor Hybrid Structures}

\author{Jasmin Bedow$^{1}$, Eric Mascot$^{2}$, Themba Hodge$^{2}$, Stephan Rachel$^{2}$ and Dirk K. Morr$^{1}$}
\affiliation{$^{1}$ Department of Physics, University of Illinois at Chicago, Chicago, IL 60607, USA\\
$^{2}$ School of Physics, University of Melbourne, Parkville, VIC 3010, Australia}

\date{\today}

\maketitle
%%%%%%%%%%%%%%%%%%%%%%%%%%%%%%%%%%%%%%%%%%%%%%%%%%%%%%%%%%
%%%%%%%%%%%%%%%%%%%%%%%%%%%%%%%%%%%%%%%%%%%%%%%%%%%%%%%%%% }
{\bf The creation of topological quantum gates using Majorana zero modes -- an outstanding problem in the field of topological quantum computing -- relies on our ability to control the braiding process of these particles in time and space. Here, we demonstrate the successful implementation of topologically protected  $\sqrt{\sigma_z}$-, $\sigma_z$- and $\sigma_x$-quantum gates using Majorana zero modes in two-dimensional magnet-superconductor hybrid structures. We propose a braiding protocol that makes use of recent advances in the ability to control the spin of individual atoms using electron-spin-resonance techniques.  We visualize the braiding process in time and space by computing the non-equilibrium local density of states, which is proportional to the time-dependent differential conductance measured in scanning tunneling spectroscopy experiments. \\}

{\it Introduction.~}
Majorana zero modes (MZMs) that are realized in topological superconductors provide an intriguing platform for the implementation of fault-tolerant quantum computing \cite{Nayak2008}. An important step on this path is the realization of topological quantum gates, for which various protocols have been proposed \cite{Alicea2011,Halperin2012,Kraus2013,Amorim2015,Aasen2016,Sekania2017,Karzig2017,Harper2019,Tutschku2020,Sanno2021,Tanaka2022}.  Common to all of these proposals is that they require atomic scale control of the electronic or magnetic structure, in terms of either the local chemical potential \cite{Alicea2011,Halperin2012,Sekania2017,Harper2019,Tutschku2020,Tanaka2022}, coupling constants \cite{Kraus2013,Amorim2015,Zhou2022}, superconducting phases \cite{Sanno2021} or magnetic fields \cite{Hyart2013,Li2016}. How this control can be achieved experimentally, and how a successful implementation of the gate operation can be visualized, have remained some of the most challenging open problems in the field.

In this article, we demonstrate that magnet-superconductor hybrid (MSH) systems, consisting of networks of magnetic adatoms placed on the surface of $s$-wave superconductors, provide a versatile platform for the implementation of topological quantum gates using Majorana zero modes. Such systems can be built using atomic manipulation techniques \cite{Kim2018a}, and possess the great advantage that their local, atomic scale magnetic structure can be manipulated using a combination of electron-spin resonance and scanning tunneling microscopy (ESR-STM) techniques \cite{Yang2019,Wang2021,Phark2022}. The latter, in turn, allows one in general to switch the system between trivial and topological phases, thus facilitating the braiding of MZMs and enabling the realization of topological $\sqrt{\sigma_z}$-, $\sigma_z$- and $\sigma_x$-quantum gates. We show that the spatial exchange of MZMs, and the gate operation in its entirety can be visualized through the time-, energy-, and spatially resolved non-equilibrium density of states \cite{Bedow2022}, which can be experimentally imaged via the time-dependent differential conductance, $dI/dV$, measured in scanning tunneling spectroscopy (STS) \cite{Balatsky2006,vanHouselt2010}. We demonstrate that the successful implementation of topological quantum gates can only be achieved by understanding the interplay between gate architecture and size, coherence length of the underlying superconducting phase and characteristic time of the gate operation.

\begin{figure}
  \centering
  \includegraphics[width=8.5cm]{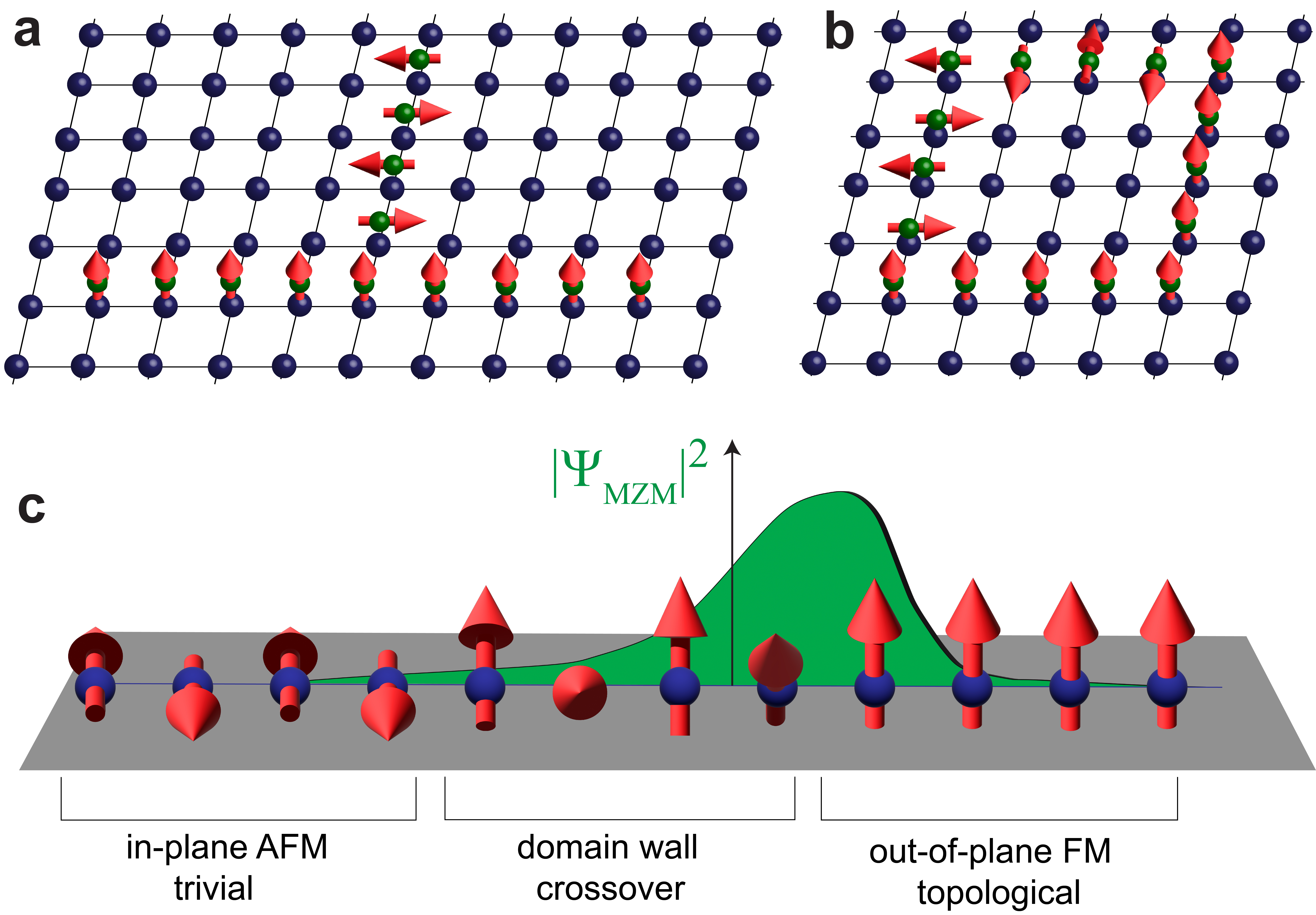}
  \caption{{\bf Topological tuning in MSH networks} Schematic representation of {\bf a} a T-structure MSH system, and {\bf b} a single loop MSH system. {\bf c} A 1D MSH network can be tuned locally between a topological and trivial phase by changing the local magnetic structure from an out-of-plane ferromagnetic to an in-plane antiferromagnetic alignment. The MZM is localized at the end of the network's topological region.
  }
  \label{fig:Fig1}
\end{figure}

{\it Theoretical Methods.~}
To simulate the braiding of MZMs, and the implementation of topological quantum gates, we consider MSH structures consisting of one-dimensional networks of magnetic adatoms placed on the surface of a two-dimensional (2D) $s$-wave superconductor (see Figs.~\ref{fig:Fig1}{\bf a},{\bf b}), described by the Hamiltonian
\begin{align}
\mathcal{H} =& \; -t_e \sum_{{\bf r}, {\bf r}', \alpha} c^\dagger_{{\bf r}, \alpha} c_{{\bf r}', \alpha} - \mu \sum_{{\bf r}, \alpha} c^\dagger_{{\bf r}, \alpha} c_{{\bf r}, \alpha} \nonumber \\
     &+ \mathrm{i} \alpha \sum_{{\bf r}, {\bm \delta }, \alpha, \beta} c^\dagger_{{\bf r}, \alpha}  \left({\bm \delta} \times \boldsymbol{\sigma} \right)^z_{\alpha, \beta}  c_{{\bf r} + {\bm \delta}, \beta} \nonumber \\
    & + \Delta \sum_{{\bf r}} \left( c^\dagger_{{\bf r}, \uparrow} c^\dagger_{{\bf r}, \downarrow} + c_{{\bf r}, \downarrow} c_{{\bf r}, \uparrow} \right) \nonumber \\
    &+  {\sum_{{\bf R} , \alpha, \beta}} c^\dagger_{{\bf R}, \alpha} \left[J_{\bf R} {\bf S}_{\bf R}(t) \cdot \boldsymbol{\sigma} \right]_{\alpha,\beta} c_{{\bf R}, \beta} \; .
    \label{eq:H}
\end{align}
Here, the operator $c^\dagger_{{\bf r}, \alpha}$ creates an electron with spin $\alpha$ at site ${\bf r}$, $t_e$ is the nearest-neighbor hopping amplitude on a 2D square lattice, $\mu$ is the chemical potential, $\alpha$ is the Rashba spin-orbit coupling between nearest-neighbor sites ${\bf r}$ and ${\bf r} +{\bm \delta}$, and $\Delta$ is the $s$-wave superconducting order parameter. The last term in Eq.(\ref{eq:H}) describes the coupling between the magnetic adatoms with spin ${\bf S}_{\bf R} (t)$ of magnitude $S$ at site ${\bf R}$ and time $t$ and the conduction electrons, with exchange coupling $J$.  Due to the hard superconducting gap, which suppresses Kondo screening \cite{Balatsky2006,Heinrich2018}, we can consider the spins of the magnetic adatoms to be classical in nature. Finally, we assume no direct exchange coupling between magnetic adatoms, which can be achieved by considering sparser networks with greater distances between adatoms, which nevertheless can be topological (see discussion below).

We choose parameters such that the networks (see Figs.~\ref{fig:Fig1}{\bf a},{\bf b}) are topological superconductors when the magnetic adatoms are aligned ferromagnetically out-of plane, but are trivial (gapped) superconductors when the moments are aligned antiferromagnetically in-plane, as schematically shown in Fig.~\ref{fig:Fig1}{\bf c}. Thus, the topological nature of these networks can be changed locally through a position-dependent rotation of magnetic moments between in- and out-of-plane, which, in turn, allows us to move MZMs through the network as they are localized at the end of the topological regions (see Fig.~\ref{fig:Fig1}{\bf c}).
Such a local control to rotate individual magnetic moments in assemblies of magnetic adatoms was recently demonstrated using ESR-STM techniques \cite{Yang2019,Wang2021,Phark2022}. Moreover, to rotate magnetic moments in opposite direction (as required to create an antiferromagnetic in-plane alignment) can be achieved by using different types of magnetic adatoms, or by changing the local magnetic structure \cite{Wang2021,Phark2022}. Below, we demonstrate that we can create topological MSH networks even for two different types of magnetic adatoms, resulting in different values of $JS$, thus opening a path to switching the magnetic structure locally between out-of-plane ferromagnetic and in-plane antiferromagnetic.

\begin{figure*}
  \centering
  \includegraphics[width=17cm]{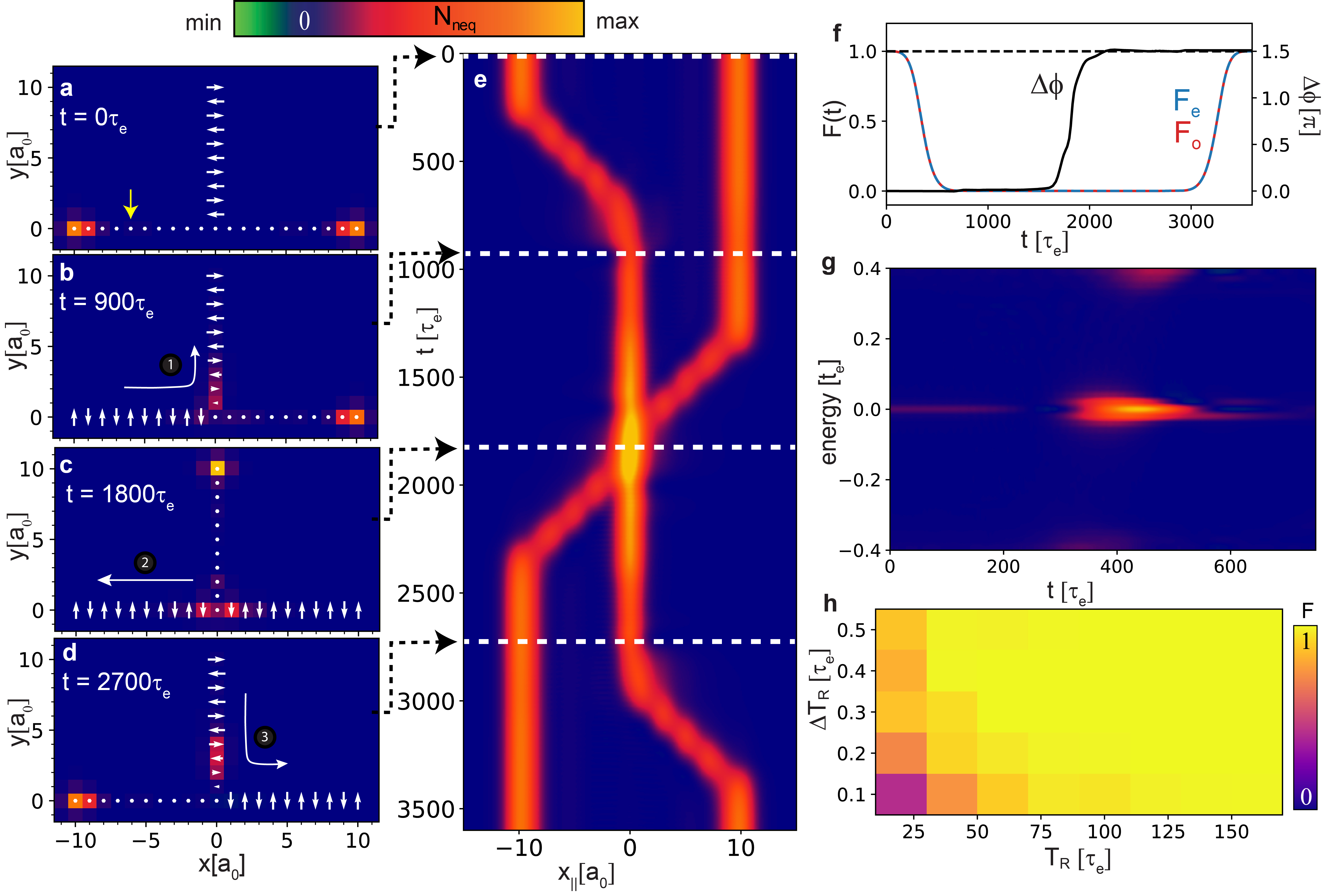}
  \caption{{\bf Implementation of a $\sqrt{\sigma_z}$-gate in an MSH T-structure}. {\bf a}-{\bf d} Spatial plot of the zero-energy $N_\text{neq}$ during consecutive times in the gate operation. White arrows (dots) indicate antiferromagnetic in-plane (ferromagnetic out-of-plane) alignment of the magnetic moments. {\bf e} Majorana world-lines obtained from a projection of the zero-energy $N_\text{neq}$ onto the real space $x$-axis. {\bf f} Time-dependence of the fidelity $F_{e,o}(t)$ for the even and odd parity states and of the geometric phase difference $\Delta \phi$. {\bf h} Time- and energy-dependent $N_\text{neq}$ at a site of the MSH network (see yellow arrow in {\bf a}). Parameters are $(\mu, \alpha, \Delta, JS) = (-3.993, 0.9, 2.4, 5.2) t_e$ with a difference of $\Delta J = 0.26 t_e$ in the magnetic coupling between alternating sites on the T-structure, $(T_\text{R}, \Delta T_\text{R}) = (500, 50) \tau_e$ and $\Gamma = 0.01 t_e$ for the inverse quasi-particle lifetime. These parameters, resulting in a topological superconducting gap of $\Delta_t \approx 0.4t_e$, were chosen in order to minimize (i) the localization length of the MZMs along the network, and (ii) thus the hybridization between the MZMs.}
  \label{fig:Fig2}
\end{figure*}

Atomic scale and time-resolved insight into the dynamics of gate operations can be gained via the time-dependent and spatially resolved differential conductance, $dI(V,{\bf r},t)/dV$, measured in scanning tunneling spectroscopy experiments \cite{Balatsky2006,vanHouselt2010}. We previously showed that, similar to the equilibrium case, $dI(V,{\bf r},t)/dV$ is proportional to the local non-equilibrium density of states $N_\mathrm{neq} (\omega=eV, {\bf r}, t) = - \frac{1}{\pi} \mathrm{Im}\left[G^r (\omega, {\bf r}, t) \right]$ \cite{Bedow2022}. Here, the retarded Green's function matrix $\hat{G}^r$ is obtained by solving the differential equation \cite{Bedow2022}
\begin{equation}
    \left[ \mathrm{i} \frac{d}{dt} +  \omega  + \mathrm{i} \Gamma - {\hat H}(t)\right] {\hat G}^r \left(t, \omega \right) = {\hat 1} \; ,
\end{equation}
with the detailed time dependence of the gate operation being encoded in the time-dependent matrix form $\hat{H}$ of the Hamiltonian in Eq.(\ref{eq:H}).  The rotation of the magnetic moments is characterized by two time scales:  the rotation time $T_\text{R}$ to rotate a single moment by $\pi/2$ between in- and out-of-plane alignment, and the delay time $\Delta T_\text{R}$ between the start of rotations on neighboring sites. Note that below all times are given in units of $\tau_\textrm{e} = \hbar /t_\textrm{e}$ which implies that for typical values of $t_\textrm{e}$ of a few hundred meV, $\tau_\textrm{e}$ is of the order of a few femtoseconds.

To ascertain the adiabaticity of the gate process, we compute the time-dependent fidelity
\begin{align}
  F_i(t) &= \left| \braket{\Psi_i(t) | \Psi_i(t_0) } \right|\ .
\end{align}
 $(i=e,o)$ of the even ($\ket{\Psi_e(t)}$) and odd-parity ($\ket{\Psi_o(t)}$) many-body wave-functions \cite{Alicea2011,Shi2017,Sekania2017}
 (for details, see Supplementary Section I).
Finally, to demonstrate the fractional statistics
of MZMs, one computes the time-dependent geometric phase, $\phi_{i}(t) (i=e,o)$, of the even and odd-parity ground state wave-function using the gauge- and parametrization-invariant functional \cite{Samuel1988, Mukunda1993}
\begin{equation}
    \phi_{i}(t) = \mathrm{arg} \braket{\Psi_{i} (t_0) | \Psi_{i} (t)} - \mathrm{Im} \int_{t_0}^t \braket{\Psi_i (t') | \dot{\Psi}_i (t')} \mathrm{d}t' \; ,
    \label{eq:geo-phase}
\end{equation}
where the exchange of two MZMs leads to a change of $\Delta\phi=\phi_{e}-\phi_{o}$ by an odd multiple of $\pi/2$ \cite{Alicea2011,Cheng2011,Sekania2017}.\\

\begin{figure*}
  \centering
  \includegraphics[width=17cm]{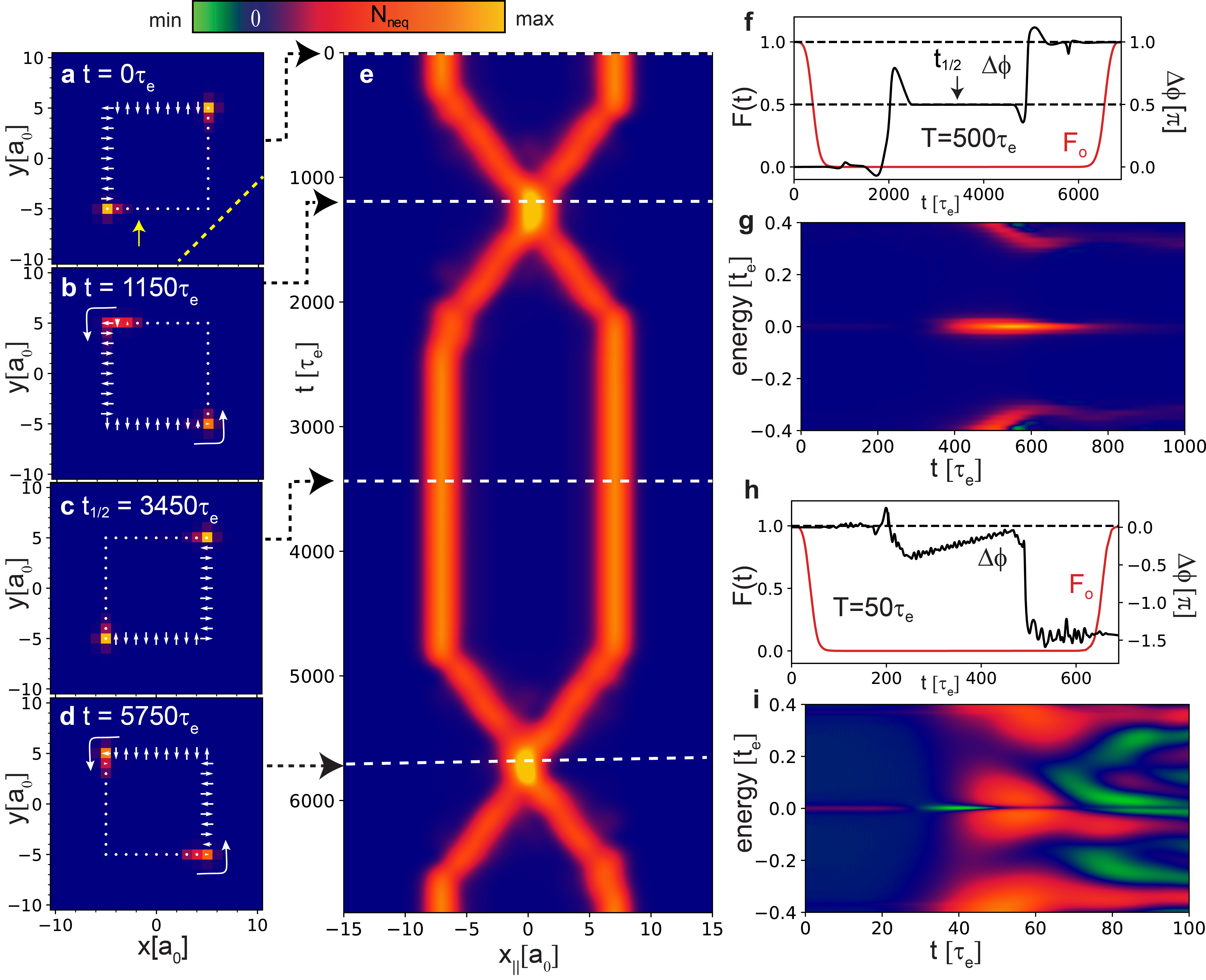}
  \caption{{\bf Implementation of a $\sigma_z$-gate in a loop MSH structure}.  {\bf a}-{\bf d} Spatial plot of the zero-energy $N_\text{neq}$ during consecutive times in the gate operation. {\bf e} Majorana world lines, as obtained from a projection of the zero-energy $N_\text{neq}$ onto the dashed yellow line in {\bf a}.  {\bf f}  Fidelity of the odd parity many-body wave-function and geometric phase as a function of time. {\bf g} Time- and energy- dependent $N_\text{neq}$ at a site in the MSH network (see yellow arrow in {\bf a}). For all results $(T_\text{R}, \Delta T_\text{R}) = (500, 100) \tau_e$. {\bf h, i} Same as {\bf f, g} but for a 10 times faster gate operation with $(T_\text{R}, \Delta T_\text{R}) = (50, 10) \tau_e$. Parameters are $(\mu, \alpha, \Delta, JS) = (-3.993, 0.9, 2.4, 5.2) t_e$, and $\Gamma = 0.01 t_e$.}
  \label{fig:Fig3}
\end{figure*}
{\it Results.~} Two basic gate architectures have previously been proposed to implement topological quantum gates in 1D systems -- a T-structure \cite{Alicea2011} (Fig.~\ref{fig:Fig1}{\bf a}) or a loop structure \cite{Li2016} (Fig.~\ref{fig:Fig1}{\bf b}) --, which can be built using atomic manipulation techniques on the surface of a 2D superconductor \cite{Kim2018a}. We demonstrate the feasibility of implementing gate protocols in both of these architectures, beginning with a $\sqrt{\sigma_z}$-gate -- realizing the exchange of two MZMs -- in a T-structure network. To demonstrate the fractional statistics of MZMs, it is necessary for the gate process to be adiabatic, thus avoiding excitations between the MZMs and bulk states. To this end, we choose a rotation time $T_\text{R} \gg \hbar/\Delta_t$, where $\Delta_t$ is the topological gap in the system (the time-dependent gate protocol is given in Supplementary Section II), and present in Figs.~\ref{fig:Fig2}{\bf a}-{\bf d}, the resulting zero-energy $N_\text{neq}$ at successive times during the gate process together with the magnetic structure, shown as white arrows (the full time dependence of the entire gate process is shown in Supplementary Movie 1). At the initial time $t=0$, two MZMs are localized at the ends of the topological horizontal segment of the T-structure (cf.~Fig.~\ref{fig:Fig1}{\bf a} and Fig.~\ref{fig:Fig2}{\bf a}), while the vertical segment is trivial.  The exchange of the two MZMs is then facilitated in three steps, as schematically shown in Figs.~\ref{fig:Fig2}{\bf b}-{\bf d}. A comparison between the spatial form of $N_\text{neq}$ and that of the magnetic structure, reveals as expected, that the spatial location of the MZMs follows the boundary between the network's topological and trivial regions, as schematically shown in Fig.~\ref{fig:Fig1}{\bf c}. Once the braiding process is completed, $N_\text{neq}$ exhibits the same spatial structure as in the initial state (see Fig.~\ref{fig:Fig2}{\bf a}). The resulting Majorana world-lines (see Fig.~\ref{fig:Fig2}{\bf e}), obtained by projecting the zero-energy $N_\text{neq}$ onto the real space $x$-axis  visualize the entire gate process in time and space (a 3D rendering of the world-lines is presented in Supplementary Movie 2).

\begin{figure*}
  \centering
  \includegraphics[width=17cm]{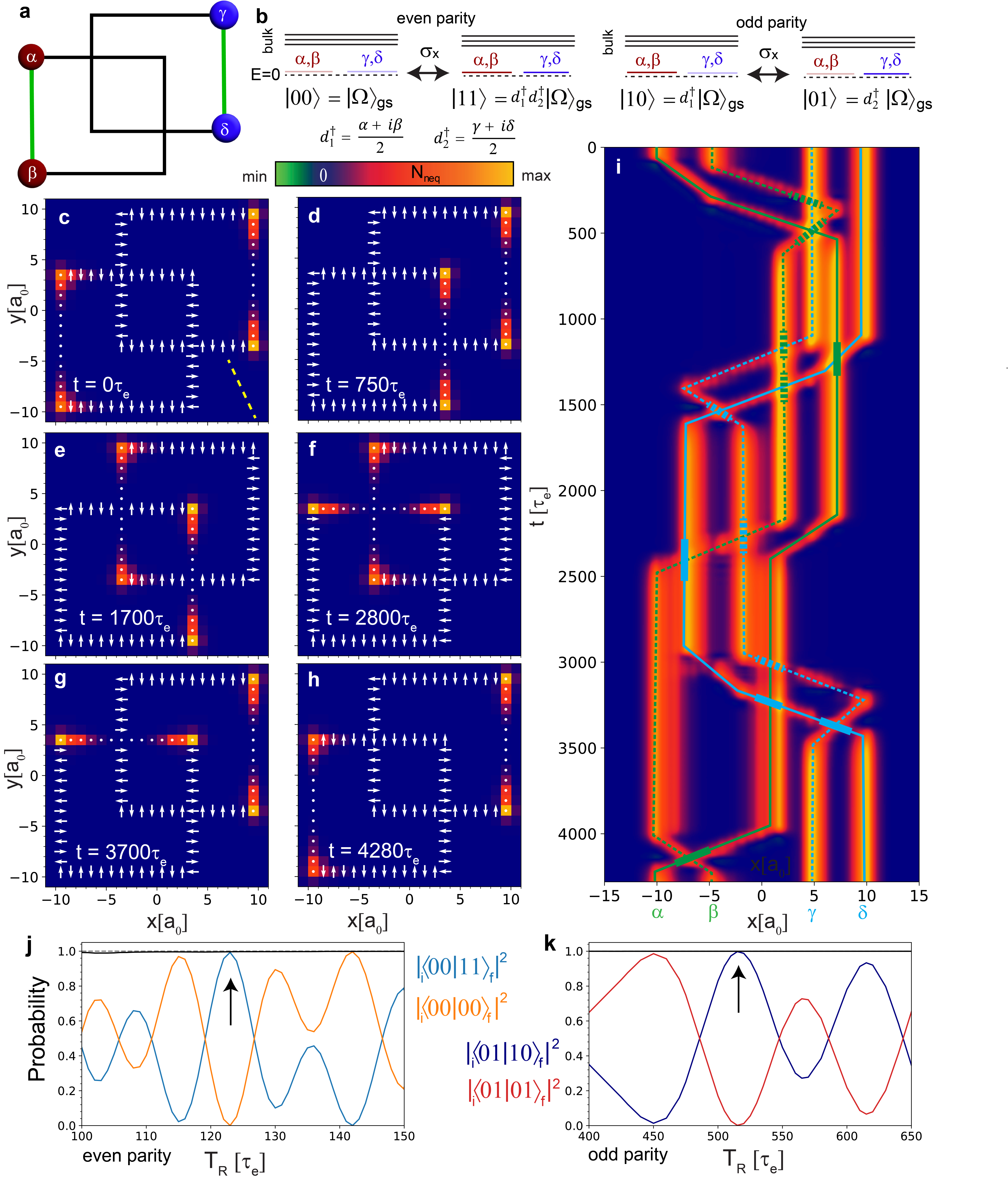}
  \caption{{\bf Implementation of a $\sigma_x$-gate in a double-loop  MSH structure}. {\bf a} Schematic picture of the $\sigma_x$-gate in an MSH system, consisting of two intersecting loops of magnetic adatoms, and two pairs of MZM. {\bf b} Effect of the $\sigma_x$-gate on the even and odd parity many-body wave-functions.  {\bf c} - {\bf h} Spatial plot of the zero-energy $N_\text{neq}$ at successive times during the gate process. {\bf i} Majorana world lines, as obtained from a projection of zero-energy $N_\text{neq}$ onto the dashed yellow line in {\bf c}. Transition amplitudes within the {\bf j} even parity sector with $\mu = -3.5 t_e$, $\Delta T_\text{R} = 0.2 T_\text{R}$, and {\bf k} odd parity sector for $\mu = -3.45805 t_e$, $\Delta T_\text{R} = 0.1 T_\text{R}$.  For all other panels, parameters are $(\mu, \alpha, \Delta, JS) = (-3.993,0.9, 2.4, 5.2) t_e$, $(T_\text{R},\Delta T_\text{R}) = (100, 20) \tau_e$, and $\Gamma = 0.01 t_e$.}
  \label{fig:Fig4}
\end{figure*}

The time-dependence of $F_{e,o}(t)$ during the entire gate process is shown in Fig.~\ref{fig:Fig2}{\bf f}. Due to the evolving magnetic structure, $F_{e,o}(t)$ deviates from unity after the start of the gate process, quickly reaching nearly zero due to the approximate orthogonality between the initial many-body states at $t=0$ and that at $t$. However, at the end of the gate process, when the initial magnetic structure is reestablished, the fidelity returns to near unity (with a deviation of less than $0.005$), demonstrating the (near complete) adiabaticity of the braiding process. As a result, the geometric phase $\Delta \phi$ reaches $3\pi/2$ at the end of the process (see Fig.~\ref{fig:Fig2}{\bf f}), establishing the fractional statistics of the MZMs.
The adiabaticity of the process is also reflected in the energy- and time-resolved $N_\text{neq}$ at a given site in the system, (see Fig.~\ref{fig:Fig2}{\bf g}) which demonstrates that as the MZM moves through a site, it remains well separated in energy from the bulk states, thus ensuring a fidelity of unity. Finally, in Fig.~\ref{fig:Fig2}{\bf h}, we present the fidelity as a function of rotation time $T_\text{R}$ and delay time $\Delta T_\text{R}$, which defines the time regime necessary to observe an adiabatic process.

We next implement a $\sigma_z$-gate, using the MSH loop structure of Fig.~\ref{fig:Fig1}{\bf b}. In the initial ($t=0$) configuration, the MZMs are localized in the upper right and lower left corners of the loop (see Fig.~\ref{fig:Fig3}{\bf a}), at the ends of the topological (ferromagnetic) segment in the loop's lower right half. We realize a $\sigma_z$-gate by moving the MZMs once around the entire loop structure (the time-dependent gate protocol is given in Supplementary Section II). The resulting zero-energy $N_\text{neq}$ together with the corresponding magnetic configuration is shown in Figs.~\ref{fig:Fig3}{\bf a}-{\bf d} for consecutive times during the gate operation (the full time dependence of $N_\text{neq}$ during the entire gate process is shown in Supplementary Movie 3). After the end of the gate process, $N_\text{neq}$ exhibits the same spatial structure as in the initial configuration (see Fig.~\ref{fig:Fig3}{\bf a}). The MZMs' world-lines (see Fig.~\ref{fig:Fig3}{\bf e}), shown as a projection of the zero-energy $N_\text{neq}$ onto the diagonal axis (see dashed yellow line in Fig.~\ref{fig:Fig3}{\bf a}), reveal the double exchange of the MZMs in space and time (a 3D rendering of the world-lines is presented in Supplementary Movie 4). The adiabaticity of the process, as demonstrated by $F_{o}$ reaching unity at the end of the gate process (see Fig.~\ref{fig:Fig3}{\bf f}), then implies a change in the geometric phase of $\Delta \phi = \pi$. Note that after half of the gate operation at time $t_{1/2}$, the two MZMs have been exchanged, which realizes a $\sqrt{\sigma_z}$-gate, similar to the case of Fig.~\ref{fig:Fig2}. However, while the geometric phase $\Delta \phi$ at this point has as expected changed by $\pi/2$ (Fig.~\ref{fig:Fig3}{\bf f}), the corresponding fidelity is zero. The latter, however, is not a reflection of the non-adiabaticity of the process, but of the fact that the initial spin configuration (Fig.~\ref{fig:Fig3}{\bf a}), and that at time $t_{1/2}$ differ significantly (Fig.~\ref{fig:Fig3}{\bf c}). Moreover, the adiabaticity of the gate process is again reflected in the energy- and time-dependence of $N_\text{neq}$ at a site in the loop (see Fig.~\ref{fig:Fig3}{\bf g}), which demonstrates that the MZM and the bulk states remain well separated in energy during the gate process. To contrast this, we consider a 10-times faster gate operation: while in this case, the fidelity $F_{o}=0.98$ is only slightly reduced from unity  (Fig.~\ref{fig:Fig3}{\bf h}), the geometric phase deviates already strongly from the expected value of $\pm \pi$, clearly revealing the breakdown of adiabaticity. This is further confirmed by a plot of $N_\text{neq}$ (see Fig.~\ref{fig:Fig3}{\bf i}) that reveals a strong overlap in energy between the MZM and bulk states. We thus conclude that in addition to the fidelity, $N_\text{neq}$ reflects the adiabaticity, or lack thereof, of the gate operation, thus providing an experimentally measurable signature of an adiabatic gate process.

Finally, to implement a one-qubit $\sigma_x$-gate we consider an MSH system consisting of two intersecting loops of magnetic adatoms, as schematically shown in Fig.~\ref{fig:Fig4}{\bf a}. In each loop, a pair of MZMs, labelled $\alpha,\beta$ and $\gamma,\delta$, are localized at the ends of their respective ferromagnetic, and hence topological, segments (shown in green).  The many-body wave-functions in the even and odd parity sector are built as schematically shown in Fig.~\ref{fig:Fig4}{\bf b} with $|\Omega \rangle_{\rm gs}$ being the many-body ground state wave-function. Due to the finite hybridization of the MZMs within each pair, their respective energies are small, but non-zero (see Fig.~\ref{fig:Fig4}{\bf b}).
The $\sigma_x$-gate operation, transforming the two states within each parity sector into one another, is implemented as shown in Figs.~\ref{fig:Fig4}{\bf c} - {\bf h}, where we present spatial plots of the zero-energy $N_\text{neq}$ for consecutive times  during the gate process (the time-dependent gate protocol is given in Supplementary Section II, and the full time dependence of $N_\text{neq}$ is shown in supplementary Movie 5). The resulting Majorana world-lines (Fig.~\ref{fig:Fig4}{\bf i}), obtained by projecting the zero-energy $N_\text{neq}$ onto the real space axis shown as a dashed yellow line in  Fig.~\ref{fig:Fig4}{\bf c} visualize the gate operation, and in particular the double exchange of the $\beta$ and $\gamma$ MZMs in time and space (a 3D rendering of the world-lines is presented in Supplementary Movie 6). To demonstrate that the braiding of MZMs shown in Figs.~\ref{fig:Fig4}{\bf c} - {\bf h} indeed constitutes a $\sigma_x$-gate, we compute the transition probabilities in the even and odd parity sector via $p_e = |_i\langle 00|11\rangle_f|^2$ and $p_o = |_i\langle 01|10\rangle_f|^2$ where the subscripts $i,f$ denote the initial ($t=0$) and final time ($t=t_f$) many-body wave-functions. We find that $p_{e,o}$ undergoes periodic oscillations as a function of the characteristic time scale of the gate operation, given by the rotation time $T_\text{R}$, as shown in Figs.~\ref{fig:Fig4}{\bf j},{\bf k} for two different parameter sets. The maximum transition amplitudes of $p_e=0.993$  and $p_o=0.999$ (see black arrows) demonstrate that a successful realization of the $\sigma_x$-gate can be achieved for specific rotation times, $T_\text{R}$. It was previously suggested \cite{Cheng2011,Scheurer2013,Harper2019,Sanno2021} that such oscillatory behavior of $p_{e,o}$ is due to a finite energy splitting between the two many-body states within each parity sector (see Supplementary Section III). Such a splitting arises from the finite hybridization between the MZMs in systems whose size is not significantly larger than the MZM localization length; the latter being given by the superconducting coherence length, $\xi_c$, along the network direction.  While we are computationally limited to the 2D MSH system sizes shown in Fig.~\ref{fig:Fig4}, we can test this idea, and
realize larger distances between the MZMs, by implementing a $\sigma_x$-gate in a 1D T-structure (see Supplementary Section IV). While such a gate also exhibits oscillations in $p_{e,o}$, we find that their periods can be significantly increased with increasing distance between the MZMs, and hence decreasing hybridization and energy splitting between the many-body states.
We note that the oscillations in $p_{e,o}$ are not related to any transitions between the Majorana and bulk states, as they occur even when the total transition probability within the Majorana sector, e.g.,  $|_i\langle 00|00\rangle_f|^2+ |_i\langle 00|00\rangle_f|^2$ for the even parity sector, is approximately 1 (see black lines in Figs.~\ref{fig:Fig4}{\bf j},{\bf k}). While the oscillatory behavior in $p_{e,o}$ thus seems to be a universal feature of a $\sigma_x$-gate (i.e., independent of the particular gate architecture), it would be desirable for its successful experimental realization to have $p_{e,o} \approx 1$ over extended ranges of $T_\text{R}$ to  avoid the need for an experimental fine-tuning of time scales.  This could in general be achieved by building an MSH system whose size is much larger than the localization length of the MZMs, thus yielding long oscillation periods. Given the large coherence length of many $s$-wave superconductors \cite{Kittel2004}, this would likely require gate sizes of the order of hundreds of nanometers. An alternative solution could be provided by the flexibility to implement MSH gates using different gate architectures and protocols. Indeed, a $\sigma_x$-gate can not only be implemented using different architectures, such as a double-loop (see Fig.~\ref{fig:Fig4}) or $T$-architecture (see Supplementary Section IV), but also by using different gate protocols within the same (double-loop) architecture: (i) the gate protocol of Fig.~\ref{fig:Fig4}, and (ii) a gate protocol in which the $\beta$- and $\gamma$ MZMs (see Fig.~\ref{fig:Fig4}{\bf a}) are stationary, while the $\alpha$- and $\delta$-MZMs are exchanged twice via the inner loop (see Supplementary Section V).
To what extent the flexibility to build different MSH gate architectures in combination with different gate protocols can be utilized to optimize the outcome of gate operations is a question whose answer will be of utmost importance for the experimental realization of topological quantum computing.\\

{\it Discussion.~} We have demonstrated the feasibility to implement topological quantum gates in MSH systems based on ESR-STM techniques, and to visualize the gate operation, and the resulting Majorana world-lines, via the non-equilibrium density of states, $N_\text{neq}$. As the latter is proportional to the time-dependent differential conductance, $dI(V,t)/dV$, clear signatures for the adiabaticity (or the lack thereof) of the gate operation, which occur in the time- and energy-dependence of $N_\text{neq}$ (see Figs.~\ref{fig:Fig3}{\bf g} and {\bf i}), can thus also directly be probed in STS experiments. We showed that quantum gates can not only be realized in different MSH architectures [such as the $\sqrt{\sigma_z}$-, and $\sigma_z$-gates in Figs.~\ref{fig:Fig2} and \ref{fig:Fig3}, or the $\sigma_x$-gate in a double-loop (Fig.~\ref{fig:Fig4}) or $T$-architecture (see Supplementary Section IV)], but also within the same double-loop architecture using different gate protocols. Indeed, the implementation of n-qubit gates might be facilitated by generalizations of the double-loop architecture, not only because of the flexibility provided by implementing different gate protocols, but also because of the added advantage that in such architectures, any two MZMs can be directly braided without requiring the prior exchange of any intermediate MZMs. Finally, we note that though we considered dense MSH networks, in which neighboring sites are occupied by magnetic adatoms, we find that topological networks can also be created by using sparser arrangements of magnetic adatoms, occupying only every second or third site. Such sparser networks might facilitate the use of ESR-STM techniques \cite{Yang2019,Wang2021,Phark2022} to manipulate the local electronic structure. Our results thus represent the proof of concept that the combination of atomic manipulation techniques to quantum engineer MSH structures, and of ESR-STM techniques to implement gate protocols yields a versatile platform for the realization of topological quantum gates.\\

{\it Acknowledgments}
We would like to thank C.P. Lutz and R. Wiesendanger for stimulating discussions.
J.B.\ and D.K.M.\ acknowledge support by the U.\ S.\ Department of Energy, Office of Science, Basic Energy Sciences, under Award No.\ DE-FG02-05ER46225. S.R.\ acknowledges support from the Australian Research Council through Grant No.\ DP200101118.

\end{document}

\title{Implementation of Topological Quantum Gates in Magnet-Superconductor Hybrid Structures \\
Supplemental Material}

\author{Jasmin Bedow$^{1}$, Eric Mascot$^{2}$, Themba Hodge$^{2}$, Stephan Rachel$^{2}$ and Dirk K. Morr$^{1}$}
\affiliation{$^{1}$ Department of Physics, University of Illinois at Chicago, Chicago, IL 60607, USA\\
$^{2}$ School of Physics, University of Melbourne, Parkville, VIC 3010, Australia}

\maketitle

\section*{Supplementary Section 1: Theoretical Formalism}

\subsection{Construction of ground-state wave-functions}

The Hamiltonian in Eq.~(1) of the main text can be recast into the Bogoliubov de Gennes (BdG) form
\begin{align}
    \mathcal{H}(t) = \frac{1}{2} \sum_{i,j}
    \begin{pmatrix} c_i^\dag & c_i \end{pmatrix}
    \underbrace{\begin{pmatrix} H_{ij}(t) & \Delta_{ij} \\[10pt] \Delta_{ji}^* & -H_{ij}^*(t) \end{pmatrix}}_{H_\text{BdG}}
    \begin{pmatrix} c_j \\[10pt] c_j^\dag \end{pmatrix},
\end{align}
with $H_\text{BdG}$ possessing a particle-hole symmetry, as reflected in $H_\text{BdG} = -\tau_x H_\text{BdG}^* \tau_x$, where $\tau_x$ is a Pauli matrix.
At $t=0$, the Bogoliubov transformation,
\begin{align}
    \begin{pmatrix} c_j \\[10pt] c_j^\dag \end{pmatrix}
    = \sum_n \begin{pmatrix}
        U_{jn} & V_{jn}^* \\[10pt]
        V_{jn} & U_{jn}^*
    \end{pmatrix}
    \begin{pmatrix} d_n \\[10pt] d_n^\dag \end{pmatrix},
\end{align}
diagonalizes the Hamiltonian as
\begin{align}
    \mathcal{H}(0) = \sum_n E_n \left(d_n^\dag d_n - \frac{1}{2}\right),
\end{align}
where \(E_n \ge 0\).
The ground state is the quasiparticle vacuum, $\ket{\Omega}$, such that $d_n \ket{\Omega} = 0$ for all $n$.
We construct the quasiparticle vacuum as a product state \cite{Ring1980,Alicea2011,Shi2017}.
This is done by annihilating all quasiparticles from the true $c$-particle vacuum.
\begin{align}
    \ket{\Omega} = \frac{1}{\sqrt{\mathcal{N}}} d_1 \dots d_N \ket{0}.
\end{align}
The normalization is given by $\mathcal{N} = |\det(V)|$.
The degenerate ground states are thus
\begin{align}
    \ket{00} &= \ket{\Omega}, \nonumber \\[3pt]
    \ket{01} &= d_2^\dagger \ket{\Omega},  \nonumber \\[3pt]
    \ket{10} &= d_1^\dagger \ket{\Omega}, \nonumber \\[3pt]
    \ket{11} &= d_1^\dagger d_2^\dagger \ket{\Omega}.
\end{align}

\subsection{Time evolution of states}

We define
\begin{align}
    d_n(t) = \mathcal{U}(t) d_n \mathcal{U}^{-1}(t)
\end{align}
where \(\mathcal{U}(t)\) is the unitary time-evolution operator,
\begin{align}
    \mathcal{U}(t) = \mathcal{T} \exp\left[
        -\frac{\mathrm{i}}{\hbar} \int_0^t dt' \mathcal{H}(t')
    \right]\ .
\end{align}
Using the time-dependent BdG equations \cite{Cheng2011,Amorim2015,Sanno2021}, the time-evolved operators are given by
\begin{align}
    \begin{pmatrix} d_n^\dagger(t) & d_n(t) \end{pmatrix}
    = \sum_i
    \begin{pmatrix} c_i^\dag & c_i\end{pmatrix}
    \begin{pmatrix}
        U_{in}(t) & V^*_{in}(t) \\[10pt]
        V_{in}(t) & U^*_{in}(t)
    \end{pmatrix},
\end{align}
where
\begin{align}
    \begin{pmatrix}
        U(t) & V^*(t) \\[10pt]
        V(t) & U^*(t)
    \end{pmatrix}
    &= \mathcal{T}\exp\left[
        -\frac{\mathrm{i}}{\hbar} \int_0^t dt' H_{\text{BdG}}(t')
    \right]
    \begin{pmatrix}
        U & V^* \\[10pt]
        V & U^*
    \end{pmatrix}.
\end{align}
We can now write the time-evolved ground states
\begin{align}
    \ket{00(t)} &= \ket{\Omega(t)}, \nonumber \\[3pt]
    \ket{01(t)} &= d_2^\dagger(t) \ket{\Omega(t)}, \nonumber \\[3pt]
    \ket{10(t)} &= d_1^\dagger(t) \ket{\Omega(t)}, \nonumber \\[3pt]
    \ket{11(t)} &= d_1^\dagger(t) d_2^\dagger(t) \ket{\Omega(t)}.
\end{align}
The time-evolved quasiparticle vacuum is given by
\begin{align}
    \ket{\Omega(t)} = \frac{e^{\mathrm{i}\alpha(t)}}{\sqrt{\mathcal{N}(t)}}
    \prod_{k} d_k(t) \ket{0}.
\end{align}
The normalization is given by $\mathcal{N}(t) = |\det(V(t))|$.
The phase $\alpha(t)$ arises from the evolution of the true vacuum. However, this phase is gauged away in our gauge-invariant formulation of physical quantities, such as the geometric phase.

\subsection{Overlaps between states}

For states $\ket{\psi}, \ket{\psi'} \in \{\ket{00}, \ket{01}, \ket{10}, \ket{11}\}$, the overlaps have the form
\begin{align}
    \braket{\psi'(0) | \psi(t)}
    = (-1)^s \frac{e^{\mathrm{i}\alpha(t)}}{\sqrt{\mathcal{N} \mathcal{N}(t)}}
    \braket{0|
        \prod_{k} d_k^\dagger
        (d_1)^{n_1'} (d_2)^{n_2'}
        (d_1^\dagger(t))^{n_1} (d_2^\dagger(t))^{n_2}
        \prod_{k} d_k(t)
    |0}.
\end{align}
The minus sign is due to reversing the order of the operators in $\bra{\psi'}$ and $s = (n_1'+n_2')(n_1'+n_2'-1)/2 + N(N-1)/2$.
The vacuum overlap can now be calculated using Wick's theorem \cite{Bertsch2012,Terhal2002},
\begin{align}
    \braket{\psi'(0) | \psi(t)}
    = (-1)^s \frac{e^{\mathrm{i}\alpha(t)}}{\sqrt{\mathcal{N} \mathcal{N}(t)}}
    \text{pf}(M).
    \label{eq:overlap}
\end{align}
The matrix \(M\) is an anti-symmetric matrix constructed from the contractions between operators, and $\text{pf}(\,\cdot\,)$ is the Pfaffian.
The resulting matrix is \cite{Bertsch2012,Carlsson2021,Jin2022}
\begin{align}
    M = \begin{pmatrix}
        V^T(0) U(0) &
        V^T(0) V^*(0) &
        V^T(0) U(t) &
        V^T(0) V^*(t) \\[3pt]
        &
        U^\dagger(0) V^*(0) &
        U^\dagger(0) U(t) &
        U^\dagger(0) V^*(t) \\[3pt]
        &
        &
        V^T(t) U(t) &
        V^T(t) V^*(t) \\[3pt]
        &
        &
        &
        U^\dagger(t) V^*(t) \\
    \end{pmatrix}\ .
\end{align}
Note that rows and columns corresponding to unoccupied modes must be truncated \cite{Bertsch2012}.
The lower triangle is found using anti-symmetry.
For transition probabilities, Eq.~(\ref{eq:overlap}) simplifies to
\begin{align}
    |\braket{\psi'(0) | \psi(t)}|^2
    = \frac{1}{\mathcal{N} \mathcal{N}(t)}
    |\det(M)|.
\end{align}

\section*{Supplementary Section 2: Time-dependent Gate Protocols}
In order to implement time-dependent gate protocols, we use the function
\begin{equation}
    s(t, t_0) =
    \frac{\pi}{2}
    \begin{cases}
    0 ,& t < t_0 \\
    \sin^2\left( \frac{t-t_0}{T_\text{R}}\right) ,& t_0 \leq t \leq t_0 + T_\text{R} \\
    1 ,&  t > t_0 + T_\text{R}
    \end{cases}
\end{equation}
in order to rotate a spin over a rotation period $T_\text{R}$, starting at time $t_0$. We choose this function as it allows a smooth transition from 0 to $\pi/2$ for the polar angle.
Here, we use spherical coordinates to describe each spin's orientation in space, such that
\begin{align}
{\bf S}_{\bf R}(t)  =  {
\begin{pmatrix}
    \cos(\phi({\bf R},t) \cdot \sin(\theta({\bf R},t)) \\[3pt]
    \sin(\phi({\bf R},t) \cdot \sin(\theta({\bf R},t)) \\[3pt]
    \cos(\theta({\bf R},t)) \\
\end{pmatrix}\ .
}
    \label{eq:H}
\end{align}
The azimuthal angle $\phi$ is measured with respect to the $x$-axis. Moreover,  $\Delta T_\text{R}$ is the delay time between the start of one rotation and that of the subsequent rotation on the next site, and $\Delta T_\text{wait}$ denotes a pause between steps in the gate process. Such a pause ensures that the system equilibrates between steps in the gate process.

\subsection*{A. Gate Protocol for implementing a $\sqrt{\sigma_z}$-gate in the T-Structure}
To implement a $\sqrt{\sigma_z}$-gate in the MSH T-structure of Fig.~2 in the main text, we number the sites of the magnetic adatoms from 1 to $2N_x + 1$ along the horizontal segment, where $N_x$ is the length of one leg of the T-structure and from
$2N_x + 1$ to $3N_x$ along the vertical segment. The time dependence of the azimuthal and polar angles of the spins in the network are then given by
\begin{widetext}
\begin{equation}
    (\phi_i, \theta_i(t)) = \begin{cases}
        (\frac{\pi}{2}, \quad  -(-1)^{i} s(t,(i-1) \cdot \Delta T_\text{R}) + (-1)^{i} s(t,(4N_x+2 -i) \cdot \Delta T_\text{R}),& 1 \leq i \leq N_x \\[3pt]
        (\frac{\pi}{2}, \quad  0),& i = N_x + 1\\[3pt]
        (\frac{\pi}{2}, \quad  -(-1)^{i} s(t,(4N_x + 2 - i) \cdot \Delta T_\text{R}) + (-1)^{i} s(t,(4N_x +1 + i)\cdot \Delta T_\text{R})),& N_x + 2 \leq i \leq 2N_x + 1\\[3pt]
        (0, \quad (-1)^{i} s(t,(i - N_x - 1)\cdot \Delta T_\text{R}) - (-1)^{i} s(t,(7N_x+3-i)\Delta T_\text{R})),& 2N_x + 2 \leq i \leq 3N_x + 1\\
    \end{cases}
    \label{eq:braid_switch}
    \; .
\end{equation}
\end{widetext}

\subsection*{B. Gate Protocol for implementing a $\sigma_z$-gate in the MSH loop-structure}
In the MSH loop-structure, we label the magnetic adatoms starting from the lower left corner as 1 and go counter-clockwise up until $4N_x$, where $N_x+1$ is the number of adatoms on one side of the square. We execute a $\sigma_z$ gate in two steps: In the first step, we sequentially rotate the adatoms on the lower right half of the loop into the plane starting from the lower left corner, while simultaneously rotating the upper left half out of the plane starting from the upper right corner.
In the second step, we go back to the initial configuration by reversing this process, i.e. rotating the spins in the lower right half of the loop back out of the plane starting from the lower left corner, and rotating the spins in the upper left half into the plane starting from the upper right corner.
The azimuthal and polar angles for the spins of these $4N_x$ magnetic adatoms are given as a function of time by
\begin{widetext}
\begin{equation}
    (\phi_i, \theta_i(t)) = \begin{cases}
        \left(
            \frac{\pi}{2},
            -s(t,0) + s(t,2N_x \cdot \Delta {T_\text{R}})
        \right),& i = 1 \\[3pt]
        \left(
            \frac{\pi}{2},
          (-1)^{i} s(t,(i-1) \cdot \Delta {T_\text{R}}) - (-1)^{i} s(t,(2N_x-2+i+\Delta T_\text{R}) \cdot \Delta {T_\text{R}})
        \right),& 1 < i \leq N_x \\[3pt]
        \left(
            0,
                -(-1)^{i} s(t,(i-1) \cdot \Delta {T_\text{R}})
                + (-1)^{i} s(t,(2N_x-2+i) \cdot \Delta {T_\text{R}} +\Delta T_\text{wait})
        \right),& N_x + 1 \leq i \leq 2N_x \\[3pt]
        \left(
            0,
                - s(t,T_\text{R}+2N_x\cdot\Delta T_\text{R}+\Delta T_\text{wait})
                +  s(t,T_\text{R}+(4N_x-1)\cdot\Delta T_\text{R}+\Delta T_\text{wait})
        \right), & i = 2N_x + 1\\[3pt]
        \left(
            \frac{\pi}{2},
                -(-1)^{i} s(t,(i - 2N_x -1) \cdot \Delta {T_\text{R}})
                + (-1)^{i} s(t,(i - 1) \cdot \Delta {T_\text{R}}+\Delta T_\text{wait})
        \right),& 2N_x + 1 \leq i \leq 3N_x\\[3pt]
        \left(
            0,
                (-1)^{i} s(t,(i - 2N_x -1) \cdot \Delta {T_\text{R}})
                - (-1)^{i} s(t,(i -1) \cdot \Delta {T_\text{R}} +\Delta T_\text{wait})
        \right),& 3N_x + 1 \leq i \leq 4N_x\\
    \end{cases}
    \; .
    \label{eq:sigmaz_loop}
\end{equation}
\end{widetext}

\subsection*{C. Gate Protocol for implementing a $\sigma_x$-gate in the MSH double loop-structure}

In order to implement a $\sigma_x$-gate in the MSH double loop structure shown in Fig.~3 in the main text, we label the adatom sites from 1 to $8 N_x$, where $N_x+1$ is the number of adatoms along one side of each square. The sites 1 to $4N_x$ are on the lower left square, starting from the lower left corner going counter-clockwise, and the sites $4N_x+1$ to $8N_x$ are on the upper right square, again starting at the lower left corner and going counterclockwise. The crossings of the two squares occurs at sites $N_x+1+d$ and $3N_x+1-d$ in the first square and at sites $5N_x+1-d$ and $7N_x+1+d$ in the secon square, where $d$ is both the horizontal and vertical distance from the lower left corner of the first square to the lower left corner of the second square.
The azimuthal and polar angles of the gate process are then given by
\begin{widetext}
\begin{equation}
\begin{split}
    &(\phi_i, \theta_i(t)) = \\
    &\begin{cases}
    \left(
        \frac{\pi}{2},
        s(t,N_x \cdot \Delta T_\text{R})
        - s (t , 8 N_x\cdot \Delta T_\text{R} + 4 \Delta T_\text{wait} )
    \right),& i = 1 \\[3pt]
    \left(
        \frac{\pi}{2},
        (-1)^{i} s(t,(i-1) \cdot \Delta T_\text{R})
        - (-1)^{i} s(t,(i-1+N_x) \cdot \Delta T_\text{R})
    \right), & 1 < i \leq N_x \\[3pt]
    \left(
        0,
        (-1)^{i} s(t,(i-1) \cdot \Delta T_\text{R})
        -(-1)^{i} s(t,(i+3N_x) \cdot \Delta T_\text{R} +2\Delta T_\text{wait})
    \right), & N_x < i \leq 2 N_x, i \neq N_x+1+d \\[3pt]
    \left(
        0,
        f_{Nx+1+d} (t)
    \right)
    , & i = N_x+1+d \\[3pt]
    \left(
        \frac{\pi}{2},
            s(t,2N_x \cdot \Delta T_\text{R})
            - s(t,7N_x \cdot \Delta T_\text{R}+4\Delta T_\text{wait})
    \right), & i = 2N_x \\[3pt]
    \left(
        \frac{\pi}{2},
            (-1)^{i} s(t,(i+2N_x-1) \cdot \Delta T_\text{R}+2\Delta T_\text{wait})
            -(-1)^{i} s(t,(i+5N_x) \cdot \Delta T_\text{R}+4\Delta T_\text{wait})
    \right), & 2N_x + 1 < i \leq 3 N_x, i \neq 3N_x+1-d \\[3pt]
    \left(
        \frac{\pi}{2},
        f_{3Nx+1-d} (t)
    \right), & i = 3N_x+1-d \\[3pt]
    \left(
        0,
        -(-1)^{3N_x} s(t,(5N_x) \cdot \Delta T_\text{R}+2\Delta T_\text{wait})
        +(-1)^{3N_x} s(t,8N_x \cdot \Delta T_\text{R}+4\Delta T_\text{wait})
    \right), & i = 3N_x+1 \\[3pt]
    \left(
        0,
        - (-1)^{i} s(t,(i-3N_x-1) \cdot \Delta T_\text{R})
        +(-1)^{i} s(t,(i+4N_x-1) \cdot \Delta T_\text{R}+4\Delta T_\text{wait})
    \right), & 3N_x+1 < i \leq 4N_x \\[3pt]
    \left(
        \frac{\pi}{2},
        - s(t,4 N_x \cdot \Delta T_\text{R}+ \Delta T_\text{wait})
        + s(t,(6 N_x + 1) \cdot \Delta T_\text{R} + 3\Delta T_\text{wait})
    \right), & i = 4N_x+1 \\[3pt]
    \left(
        \frac{\pi}{2},
        s(t,(i + N_x -1 ) \cdot \Delta T_\text{R}+ 3\Delta T_\text{wait})
        - s(t,(i + 2 N_x) \cdot \Delta T_\text{R} + 3\Delta T_\text{wait})
    \right), & 4N_x+1 < i \leq 5N_x, i \neq 5N_x+1-d \\[3pt]
    \left(
        0,
        - (-1)^{i} s(t,(i-3N_x ) \cdot \Delta T_\text{R}+ \Delta T_\text{wait})
        + (-1)^{i} s(t,(i + N_x -1 ) \cdot \Delta T_\text{R}+ 3\Delta T_\text{wait})
    \right), & 5N_x < i \leq 6N_x \\[3pt]
    \left(
        \frac{\pi}{2},
        (-1)^{i} s(t,(i-4N_x-1) \cdot \Delta T_\text{R}  + \Delta T_\text{wait})
        - (-1)^{i} s(t,(i-3N_x) \cdot \Delta T_\text{R}+ \Delta T_\text{wait})
    \right), & 6N_x < i \leq 7N_x \\[3pt]
    \left(
        0,
        (-1)^{i} s(t,(i-4N_x-1) \cdot \Delta T_\text{R}+ \Delta T_\text{wait})
        - (-1)^{i} s(t,(i-2N_x) \cdot \Delta T_\text{R} + 3\Delta T_\text{wait})
    \right), & 7N_x < i \leq 8N_x, i \neq 7N_x+1+d
    \end{cases}
    \label{eq:sigmax_doubleloop}
    \;.
\end{split}
\end{equation}
\end{widetext}
For the two crossing points of the squares we defined
\begin{align*}
    \begin{split}
        f_{Nx+1+d} (t) =& \;
        -(-1)^{N_x+d} s(t,(N_x+d) \cdot \Delta T_\text{R}) +(-1)^{N_x+d} s(t,(4N_x+d+1)\cdot \Delta T_\text{R}+2\Delta T_\text{wait}) \\
        &\;-(-1)^{N_x+d} s(t,(6N_x - d)\cdot \Delta T_\text{R} + 3\Delta T_\text{wait}) +(-1)^{N_x+d} s(t,(7N_x -d +1 )\cdot \Delta T_\text{R}+ 3\Delta T_\text{wait})
    \end{split} \\
    f_{3Nx+1-d} (t) =& \; -(-1)^{3N_x-d} s(t,(3N_x+d) \cdot \Delta T_\text{R} + \Delta T_\text{wait})
    +(-1)^{3N_x-d} s(t,(8N_x-d+1) \cdot \Delta T_\text{R} +4\Delta T_\text{wait})  \; .
\end{align*}

\newpage

\section*{Supplementary Section 3: Oscillations in the transition probabilities $p_{e,o}$}

The oscillatory dependence of $p_{e,o}$ on the characteristic time scale of the gate process was previously suggested \cite{Cheng2011,Scheurer2013,Harper2019,Sanno2021} to arise from a finite energy splitting between the two many-body states within each parity sector. Such a splitting originates from the finite hybridization between the MZMs in systems whose size is not significantly larger than the MZM localization length; the latter being given by the superconducting coherence length, $\xi_c$, along the network direction. In Supplementary Table~\ref{tab:tab_doublesquare} we present a series of cases where we compare the inverse energy splitting in the initial state, and the observed oscillation period. Here, we denote the oscillation period of the transition amplitude between the $\ket{00}$ and ${\ket{11}}$ states by $T_{0011}$ and the oscillation period of the transition amplitude between the $\ket{01}$ and $\ket{10}$ states by $T_{0110}$. While we find in general that a decrease in the energy splitting leads to an increase in the oscillation period, there also exist exceptions to this behavior.
\begin{table}[t]
    \centering
    \begin{tabular}{c|c|c|c|c|c}
        MZM Distance & {$\mu \; [t_e]$} & {$\frac{1}{\Delta E_{11} - \Delta E_{00}} \; [\tau_e]$} & {$T_{0011} \; [\tau_e]$}& {$\frac{1}{\Delta E_{10} - \Delta E_{01}} \; [\tau_e]$} & {$T_{0110} \; [\tau_e]$} \\[3pt]
        \toprule
        10 & -3.5   & 32.08025026 &	28.5 &	60.4664097	   & 53 \\[3pt]
        10 & -3.55	& 49.24301723 &	36.5 &	61.62151436	   & 93 \\[3pt]
        10 & -3.6	& 77.61149389 &	62	 &	152.1714373	   & 70 \\[3pt]
        10 & -3.707204	& 51.21434972 &	420	 &	14184.5468	   & 90 \\[3pt]
        14 & -3.36 & 115.7405706 &	100	 &  327.3998397    & 90 \\[3pt]
        14 & -3.470485& 737.9277107 &	200	 &  738.0488102    & 480\\[3pt]
        14 & -3.45805 & 532.7488345 &	350	 &  534083.6629    & 155
    \end{tabular}
    \caption{Initial energy differences (at $t=0$) between the many-body states in the even and odd parity sectors for the double-loop MSH structure and the oscillation periods in the transition probabilities. Parameters are $(\alpha, \Delta, J) = (0.9, 2.4, 5.2) t_e$, and $\Delta T_\text{R} = 0.1 T_\text{R}$.}
    \label{tab:tab_doublesquare}
\end{table}
The reason that the oscillation period cannot in general be directly tied to the initial inverse energy splitting is that this energy splitting itself changes significantly as a function of time during the gate process, as shown in Supplementary Fig.~\ref{fig:SIFig2} for 4 representative cases.
\begin{figure*}[h]
  \centering
  \includegraphics[width=10cm]{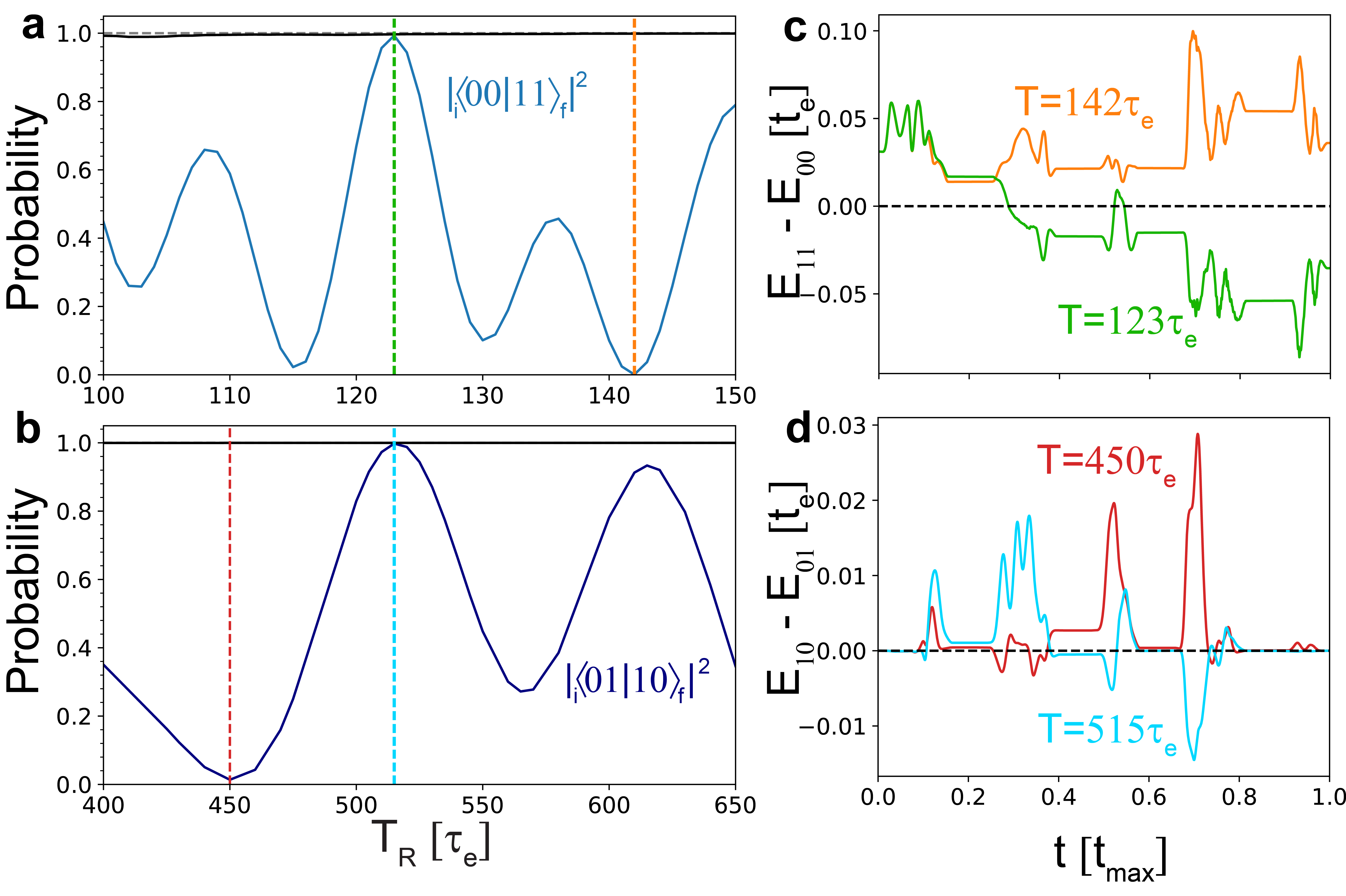}
  \caption{Transition probabilities in  {\bf a} the even parity sector for $\mu = -3.5 t_e$, and  {\bf b} the odd-parity sector with $\mu = -3.45805 t_e$ on the double-loop MSH structure of Fig.~4 of the main text. Time-evolved energy differences in the {\bf c} even and {\bf d} odd parity sector at the respective minimum and maximum in the transition probabilities from {\bf a,b}, see dashed lines.}
  \label{fig:SIFig2}
\end{figure*}

\section*{Supplementary Section 4: Implementation of a $\sigma_x$-gate in a T-structure on a 1D superconductor}

\begin{figure*}[h]
  \centering
  \includegraphics[width=17cm]{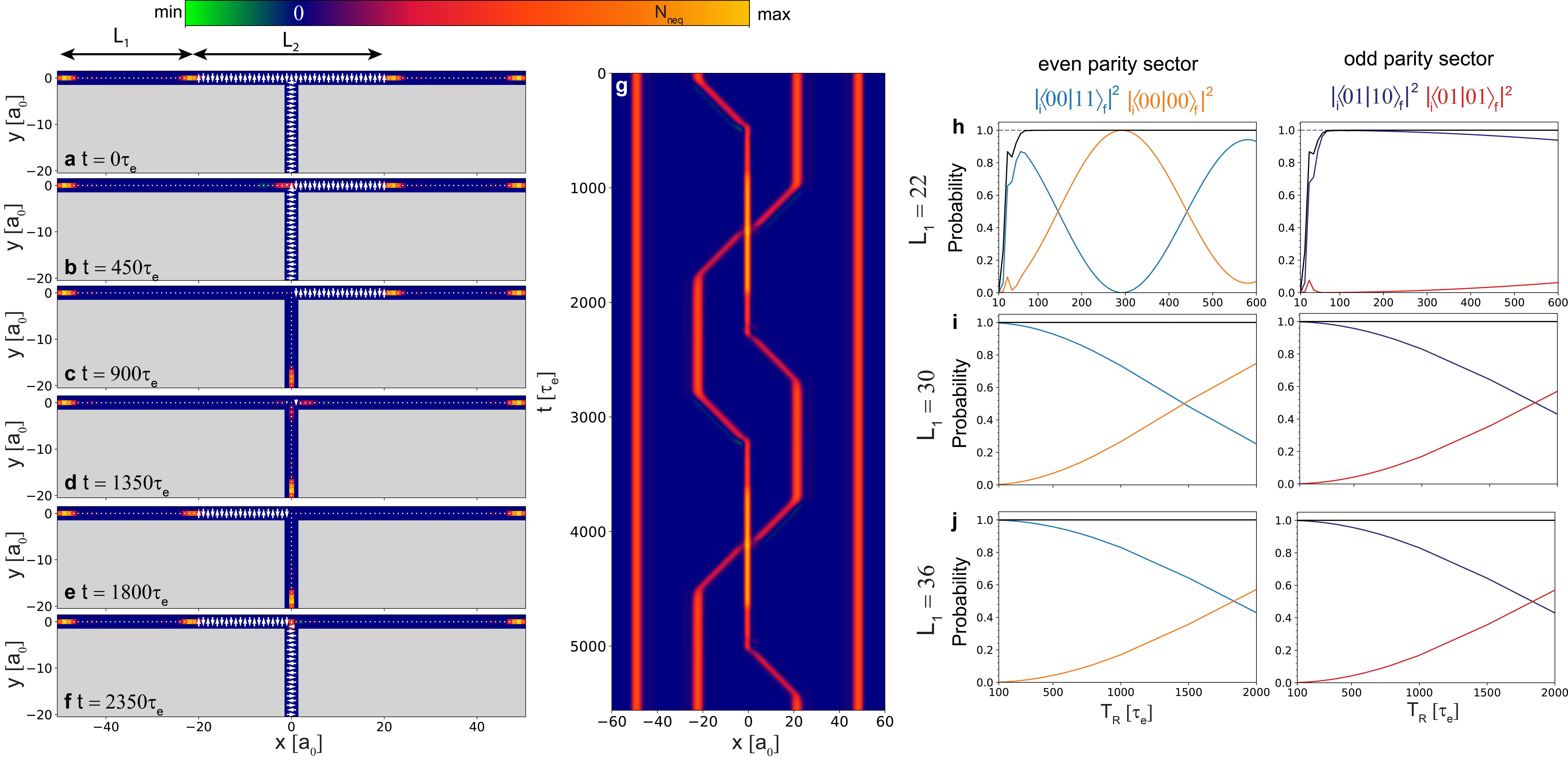}
  \caption{{\bf a} - {\bf f} Spatial plot of the zero-energy $N_{neq}$ at consecutive times during the execution of the $\sigma_x$-gate. {\bf g} Majorana world lines, as obtained from a projection of the zero-energy $N_{neq}$ onto the real-space $x$-axis. {\bf h-j} Transition probabilities within the even (left column) and odd (right column) parity sector for different distances between the two Majorana zero modes belonging to the same pair with $(\Delta T_\text{R}, \Delta T_\text{wait}) = (0.2, 1) T_\text{R}$.  Parameters used for all other panels are $(\mu, \alpha, \Delta, JS) = (-3.5,0.45, 1.2, 2.6) t_e$, $(T_\text{R},\Delta T_\text{R}, \Delta T_\text{wait}) = (100, 20, 100) \tau_e$, and $\Gamma = 0.01 t_e$.}
  \label{fig:SIFig1}
\end{figure*}
To investigate whether the oscillations  of $p_{e,o}$ seen in the implementation of the $\sigma_x$-gate using a double-loop MSH structure (see Fig.~4 of the main text), also exist in other gate architectures,  we consider a one-dimensional T-structures, as shown in Supplementary Fig.~\ref{fig:SIFig1}. For this system, the superconducting substrate possesses the same structure as the 1D network of magnetic adatoms. Thus, in contrast to the systems considered in the main text (where we considered a 1D network of magnetic adatoms on a 2D superconducting surface), this is a 1D network of magnetic adatoms on a 1D superconductor. In Supplementary Figs.~\ref{fig:SIFig1}{\bf a}-{\bf f} we present spatial plots of the zero-energy $N_{neq}$ for consecutive times during the gate process,  with the resulting Majorana world lines shown in Fig.~\ref{fig:SIFig1}{\bf g}. In Figs.~\ref{fig:SIFig1}{\bf h}-{\bf m} we present the transition probabilities for different distances $L_1$ between the MZMs within each pair (the distance $L_2=40 a_0$ is kept constant for all of these cases). Similar to the results shown in Fig.~4 of the main text, We find that the transition probabilities again exhibit an oscillatory dependence on the rotation time $T_\text{R}$.
In Supplementary Table~\ref{tab:tab_T}, we present  the inverse energy differences between the many-body states, and the observed oscillation periods for different values of $L_1$.
\begin{table}[h]
    \centering
    \begin{tabular}{c|c|c|c|c}
        MZM Distance & {$\frac{1}{\Delta E_{11} - \Delta E_{00}} \; [\tau_e]$} & {$T_{0011} \; [\tau_e]$}& {$\frac{1}{\Delta E_{10} - \Delta E_{01}} \; [\tau_e]$} & {$T_{0110} \; [\tau_e]$} \\[3pt]
        \toprule
        18 & 131.5327976 & 160	& {5.99699e+14}	& {N/A}  \\[3pt]
        20 & 1066.455819 & 1760 & {1.46927e+15}	& {N/A}  \\[3pt]
        22 & 508.4234469 & 580	& {3.8527e+14}  & {N/A}  \\[3pt]
        24 & 868.4393356 & 840	& {2.67577e+15}	& {N/A}  \\[3pt]
        30 & 6974.725568 & 5800 & {3.48739e+14}	& 7000 \\[3pt]
        36 & 86273.94678 & 7000 & {1.04533e+15}	& 7000
    \end{tabular}
    \caption{Initial energy differences in the even and odd parity sectors for the one-dimensional T-Structure with respective measured oscillation periods in the transition probabilities. Here $N/A$ implies that the oscillation period is so large that it could not be measured from the acquired numerical data (see e.g. Supplementary Fig.~\ref{fig:SIFig1}{\bf h} for $L_1 = 22$). Parameters are $(\mu, \alpha, \Delta, J) = (-3.5,0.45, 1.2, 2.6) t_e$, and $\Delta T_\text{R} = 0.2 T_\text{R}$}
    \label{tab:tab_T}
\end{table}
Similar to the results shown in Supplementary Section 3, we again find that in general an increase in the distance between the MZMs leads to an increase in the inverse energy splitting, and thus an increase in the oscillation period. These results not only suggest that the oscillations in the transition probabilities are universal and not related to any particular gate architecture, but they also provide further support for the proposal that these oscillations are directly related to the finite energy difference between the two many-body states within each parity sector.

\section*{Supplementary Section 5: Alternative Gate Protocol for a $\sigma_x$-gate in a double-loop MSH structure}

In addition to the gate protocol discussed in Fig.~4 of the main text, the double-loop MSH structure also allows for the implementation of an alternative protocol. In this case, the $\beta$ and $\gamma$ MZMs are kept stationary, and the $\alpha$ and $\delta$ MZMs are exchanged twice using the inner loop of the double-loop structure, as schematically shown in Supplementary Fig.~\ref{fig:Alt}.
\begin{figure*}
  \centering
  \includegraphics[width=15cm]{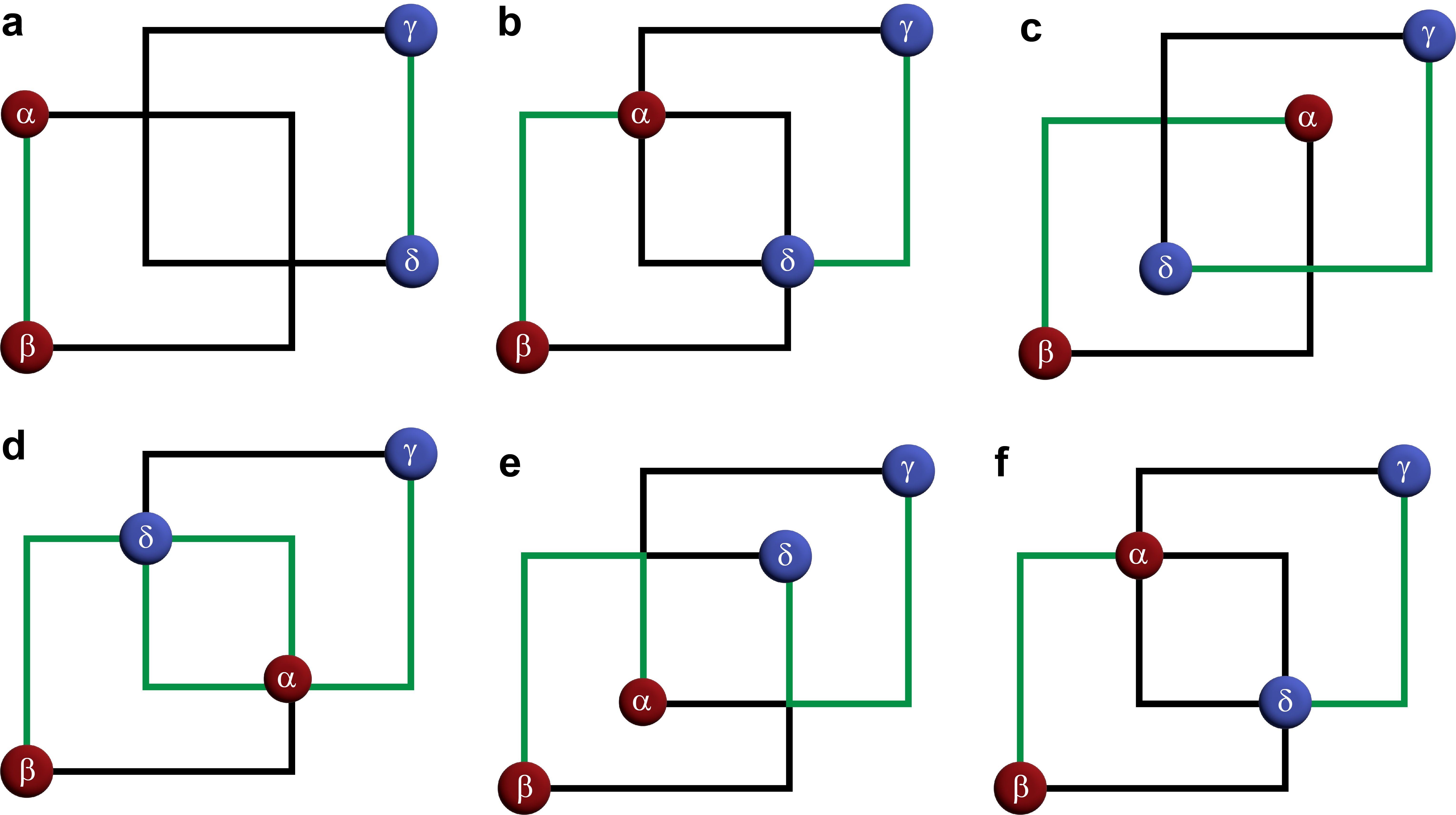}
  \caption{Schematic representation of an alternative gate protocol for the $\sigma_x$-gate. Here, only two of the four Majorana zero modes, $\alpha$ and $\delta$, are moved and exchanged twice using the inner square. We show topological regions connecting Majorana modes in green, and trivial regions in black.}
  \label{fig:Alt}
\end{figure*}
The advantages and disadvantages of this protocol and of the one shown in Fig.~4 of the main text, and in particular the question of which gate protocol is more efficient and leads to more robust transition amplitudes $p_{e,o}$ close to unity, will be studied in future work.